\documentclass[10pt,conference]{IEEEtran}
\usepackage{epsfig,rotating,setspace,latexsym,amsmath,epsf,amssymb,amsfonts,bm,theorem,cite,authblk, bbm,color, hyperref, multirow, caption, subcaption}
\IEEEoverridecommandlockouts
\allowdisplaybreaks

\title{Joint Sensing and Semantic Communications with Multi-Task Deep Learning}
\begin{document}
\author[1]{Yalin E. Sagduyu}
\author[1]{Tugba Erpek}
\author[2]{Aylin Yener}
\author[3]{Sennur Ulukus}

\affil[1]{\normalsize  Nexcepta, Gaithersburg, MD, USA}

\affil[2]{\normalsize  The Ohio State University, Columbus, OH, USA}

\affil[3]{\normalsize University of Maryland, College Park, MD, USA}
\maketitle

\begin{abstract}
This paper explores the integration of deep learning techniques for joint sensing and communications, with an extension to semantic communications. The integrated system comprises a transmitter and receiver operating over a wireless channel, subject to noise and fading. The transmitter employs a deep neural network (DNN), namely an encoder, for joint operations of source coding, channel coding, and modulation, while the receiver utilizes another DNN, namely a decoder, for joint operations of demodulation, channel decoding, and source decoding to reconstruct the data samples. The transmitted signal serves a dual purpose, supporting communication with the receiver and enabling sensing. When a target is present, the reflected signal is received, and another DNN decoder is utilized for sensing. This decoder is responsible for detecting the target's presence and determining its range. All these DNNs, including one encoder and two decoders, undergo joint training through multi-task learning, considering data and channel characteristics. This paper extends to incorporate semantic communications by introducing an additional DNN, another decoder at the receiver, operating as a task classifier. This decoder evaluates the fidelity of label classification for received signals, enhancing the integration of semantics within the communication process. The study presents results based on using the CIFAR-10 as the input data and accounting for channel effects like Additive White Gaussian Noise (AWGN) and Rayleigh fading. The results underscore the effectiveness of multi-task deep learning in achieving high-fidelity joint sensing and semantic communications.
\end{abstract}

\section{Introduction} \label{sec:Intro} 
\emph{Joint sensing and communications} (JSC), also referred to as \emph{integrated sensing and communications}, has been envisioned as a pivotal concept for next-generation communication systems such as 6G and Wi-Fi 7 by unifying information gathering (sensing) and information exchange (communication) within a common framework \cite{demirhan2022integrated, liu2022integrated, xiong2023}. This integration is instrumental for real-time decision-making, efficient resource utilization, enhancing adaptability in challenging environments, and enabling context-aware data sharing in a variety of wireless networks, such as Internet of Things \cite{cui2021integrating} and vehicle-to-infrastructure networks \cite{du2022integrated, liu2022learning}. To that end, communication signals can be dual-purposed for sensing, enabling tasks like object detection, tracking, and environmental monitoring. One application of leveraging wireless signals for sensing is WiFi sensing, where WiFi signals that are emitted as part of their standard operation are repurposed to serve as a sensing medium such that their reflected signals that bounce off objects are collected to detect objects, movements, and activities. Target sensing with JSC can provide environmental awareness and also assist with finding alternative ways of communications, e.g., using relay nodes, to avoid the objects that obstruct the line-of-sight propagation path. JSC distinguishes itself from conventional sensing methods like radar, which involve the transmission of specialized radio waves and subsequent detection of reflected signals, as opposed to utilizing existing communication signals. 

\begin{figure}[h]
	\centering
	\includegraphics[width=\columnwidth]{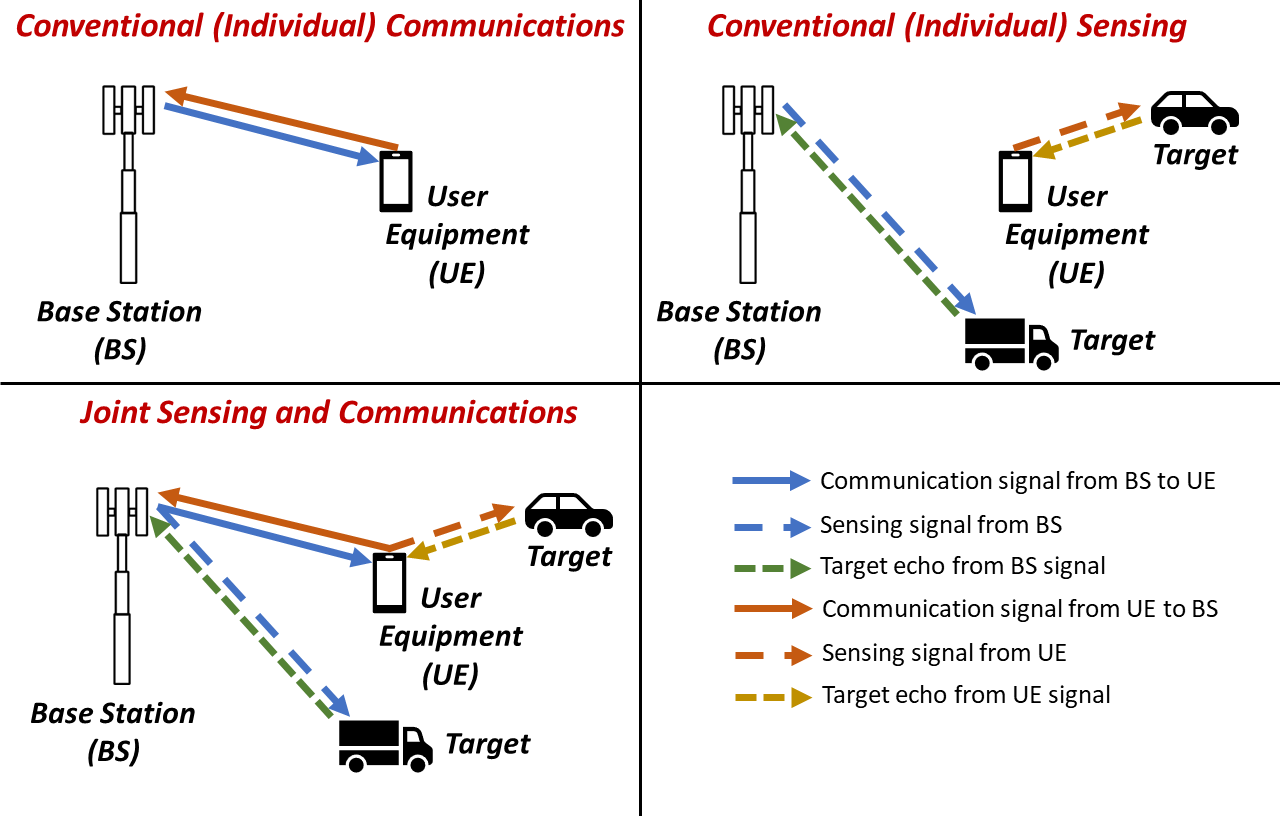}
	\caption{From individual sensing and communications to JSC.}
	\label{fig:JSCscenario}
 \vspace{0.2cm}
\end{figure}

Sensing and communications functionalities have conventionally been separate and each functionality can individually benefit from deep learning techniques due to the complexity of the underlying optimization problems and challenging data environments \cite{erpek2020deep}. Sensing and communications functionalities can be integrated in JSC by capitalizing on the broadcast nature of wireless signals. Fig.~\ref{fig:JSCscenario} illustrates the three cases of individual communications, individual sensing, and JSC. In JSC, wireless signals can be judicially designed to serve both communication and sensing functionalities, to enable applications that require real-time environmental data such as augmented reality, virtual reality or autonomous vehicles. 

In communication systems, deep neural networks (DNNs) can effectively represent (i) transmitter blocks including source coding in addition to channel coding and modulation and (ii) receiver blocks of demodulation and channel decoding followed by source decoding to reconstruct data samples such as images beyond bit/symbol representations. For example, joint source and channel coding with DNNs can outperform digital transmission concatenating JPEG and JPEG2000 compression algorithms with capacity-achieving channel codes \cite{bourtsoulatze2019deep}. Autoencoders can be trained in pairs of transmitter and receiver DNNs to jointly model transmitter and receiver functionalities of communications \cite{erpek2020deep} to optimize the fidelity of communications such as minimizing the reconstruction loss. Similarly, sensing functionality can also benefit from using DNNs for the generation of sensing waveform as well as detection of targets with the ultimate goal of maximizing the sensing accuracy. These benefits of DNNs can be reaped to perform JSC (encompassing joint operations of channel coding and modulation at the transmitter and corresponding operations at the receiver) \cite{mateosae2022}.

Compared to the single-task problems of individual sensing or communications, we utilize multi-task learning (MTL) for JSC and show that they can efficiently learn and adapt to input data, e.g., images, channel conditions, and target characteristics. In MTL, a single model is trained to perform multiple tasks of image reconstruction, target sensing, and semantic task validation simultaneously, sharing and leveraging common knowledge across tasks.

Specifically, an encoder DNN is used at the transmitter that takes data samples such as images as the input and generates wireless signals to be transmitted over a wireless channel by jointly performing source coding, channel coding, and modulation operations. The output of the encoder provides a latent representation that provides smaller dimension than the input sample, thereby reducing the number of channel uses for more efficient resource utilization. At the receiver, a decoder DNN operates on the received signals to reconstruct data samples by jointly performing demodulation, channel decoding, and source decoding operations. These two DNNs can be \emph{jointly trained for end-to-end optimization of transmitter and receiver functionalities} by adapting to input data and channel characteristics. 

To integrate sensing with communications, we assume the encoder's output signal is also used for sensing purposes. When a target is present, the signal reflects from the target, and the transmitter captures the reflected signal. We assume that the reception is directional, and the sensing receiver looks in a specific direction such that it is interested in finding if a target is in that specific angular bin (field of view) and its distance (range). To identify the presence and range of the target, the transmitter employs a third DNN, namely a decoder, for sensing. This decoder of the transmitter can be jointly trained with the encoder of the transmitter and the other decoder at the receiver to facilitate JSC by accounting for input data, channel, and potential target characteristics. For that purpose, MTL is employed where the training is performed for the combined loss for two tasks, namely reconstruction of input samples at the receiver and sensing (target detection and target position classification) at the transmitter. 

While conventional communications primarily focuses on the fidelity of information transfer by minimizing some form of reconstruction loss, there is a growing interest in \emph{preserving meaning during the information transfer}. To this end, \emph{semantic communications} has emerged as a novel communication paradigm that goes beyond the mere exchange of data or information and aims to convey the meaning, intent, and context behind the information being shared \cite{guler2018semantic}. In other words, it focuses on understanding and interpreting the significance of the transmitted messages, rather than just the raw data itself \cite{gunduz2022beyond, chaccour2022less}. Semantic communications can be formulated by training DNNs at the transmitter-receiver pair for data reconstruction while preserving semantic information \cite{xie2021deep, sagduyu2022semantic}. 

In this paper, we \emph{integrate semantic communications with JSC}. A fourth DNN, namely another decoder at the receiver, is used to validate the meaning conveyed to the receiver by performing a deep learning task (such as classification of transferred images to their labels) on received signals. This DNN is jointly trained with the other DNNs for JSC in MTL, where the task of sensing is added. This joint sensing and semantic communications (JSSC) can be converted to joint sensing and task-oriented communications by removing the data reconstruction objective and replacing semantic information validation with a general task classifier such as used in deep learning-based task (or goal)-oriented communications \cite{sagduyu2023task}.

For performance evaluation, images from CIFAR-10 dataset are used as input data samples. Both Additive White Gaussian Noise (AWGN) and Rayleigh fading channels are considered. We measure the performance in terms of peak signal-to-noise ratio (PSNR) and structural similarity index measure (SSIM) for reconstruction fidelity, sensing accuracy (both for target detection and identification of its range), and the task classification accuracy that quantifies how the meaning of information is preserved at the receiver. These measures are obtained by varying the sensing signal-to-noise ratio (SNR), namely the SNR for the signal reflected by the target and received at the transmitter), the communication SNR, namely the SNR for the signal received at the receiver, the number of potential target ranges, and the size of encoder input (namely a representative of the number of channel uses or compression level). 

As the communication baseline, we consider a conventional communications system with separate source, channel coding and modulation blocks at the transmitter and demodulation, channel and source decoding blocks at the receiver. JPEG-2000 is used for source coding, Reed-Solomon error correction code for channel coding, and 16-QAM for modulation. As the sensing baseline, we use an energy detector at the transmitter where a threshold is used to determine the presence of a target. Results show that MTL with joint training of DNNs provides an effective optimization framework for JSC as well as its extension with semantic communications, and achieves high fidelity in terms of reconstructing data samples with a greater compression ratio, preserving the meaning of information transfer and sensing targets compared to conventional communications and sensing.

The rest of the paper is organized as follows. Sec.~\ref{sec:TOC} describes MTL for JSC and its extension to semantic communications. Sec.~\ref{sec:perf} evaluates the performance.  Sec.~\ref{sec:Conclusion} concludes the paper by summarizing  the contributions and discussing future research directions. 

\begin{figure}[h]
\centering 
\includegraphics[width=\columnwidth]{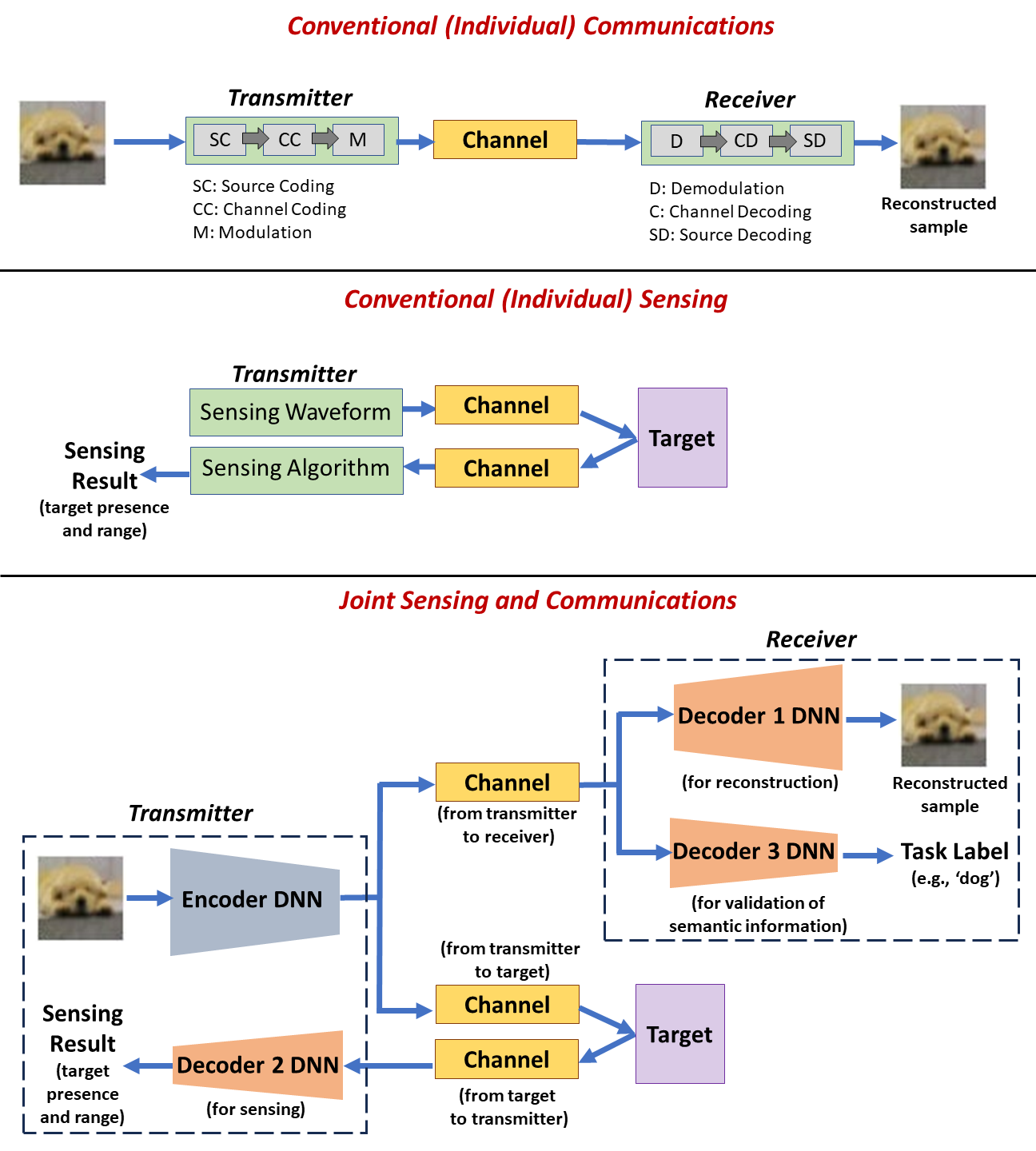  }
\caption{System model for joint sensing and (semantic) communications in comparison with conventional (individual) sensing and communications.}
\label{fig:JSCsystem}
\end{figure}

\section{Joint Sensing and Communications via Multi-task Deep Learning} \label{sec:TOC}

Fig.~\ref{fig:JSCsystem} shows the system model for JSC as an extension of system models for conventional communications and conventional sensing. DNNs are employed at the transmitter and the receiver. These DNNs are jointly trained via MTL with taking the channel characteristics into account to provide the unified functionalities of sensing and communications along with the extension to semantic communications.

\noindent \textbf{Encoder and decoders for JSSC.} The system model shown in Fig.~\ref{fig:JSCsystem} includes an encoder at the transmitter, a decoder at the receiver for data reconstruction (Decoder 1), a decoder at the transmitter for sensing of targets (Decoder 2), and a decoder at the receiver for semantic communications (Decoder 3). 
\begin{itemize}
    \item \emph{Encoder at the transmitter}: The encoder is a DNN responsible for processing input data, which includes samples like images. It performs source coding, channel coding, and modulation on the input data to generate wireless signals for transmission. The output of the encoder is a latent-space representation of the input data and serves a dual purpose: it facilitates data communication and is used for sensing. The encoder output has a smaller dimension than input samples, thereby relying on a small number of channel uses. Its role is to ensure efficient resource utilization and minimizing transmission and semantic errors. 
    \item \emph{Decoder for information reconstruction at the receiver}: Decoder 1 is a DNN at the receiver, responsible for reconstructing data samples (inputs to the encoder at the transmitter). Decoder 1 jointly performs the operations of demodulating received signals, decoding channel-encoded data, and reconstructing the original data samples with source decoding. Decoder 1 can be trained by minimizing the reconstruction loss such as the mean squared error (MSE) as the loss function. 
    \item \emph{Decoder for sensing at the transmitter}: Decoder 2, positioned at the transmitter, takes on the role of sensing. When a target is present in the environment, the signal transmitted by the encoder reflects off the target and returns to the transmitter as in radio detection and ranging (radar) systems. Assuming a stationary target, Decoder 2 is responsible for (i) either detecting a target (a binary classification problem, where the labels are the absence and presence of target) or (ii) classifying its range (that appears as signal attenuation in the reflected signal) within the environment in the presence of channel 
    effects (a multi-label classification problem, where the labels are the potential distances of the target from the transmitter in addition to the absence of target). Decoder 2 can be trained by minimizing the categorical cross-entropy as the loss function. 
    \item \emph{Decoder for semantic task classification at the receiver}: Decoder 3 is the fourth DNN in the JSC system situated at the receiver, serving the crucial role of validating the semantic content of the received data. It performs deep learning tasks that go beyond simple data reconstruction, such as classifying received signals into the labels of the corresponding data samples. By validating the semantic content, it ensures that the meaning behind the transmitted information is preserved and reliably interpreted. Decoder 3 can be trained by minimizing the categorical cross-entropy as the loss function for the underlying task classifier (such as identifying labels or the subsets of labels for the corresponding data samples that are given as input to the encoder at the transmitter). 
\end{itemize}

\begin{figure}[h]
\centering 
\includegraphics[width=\columnwidth]{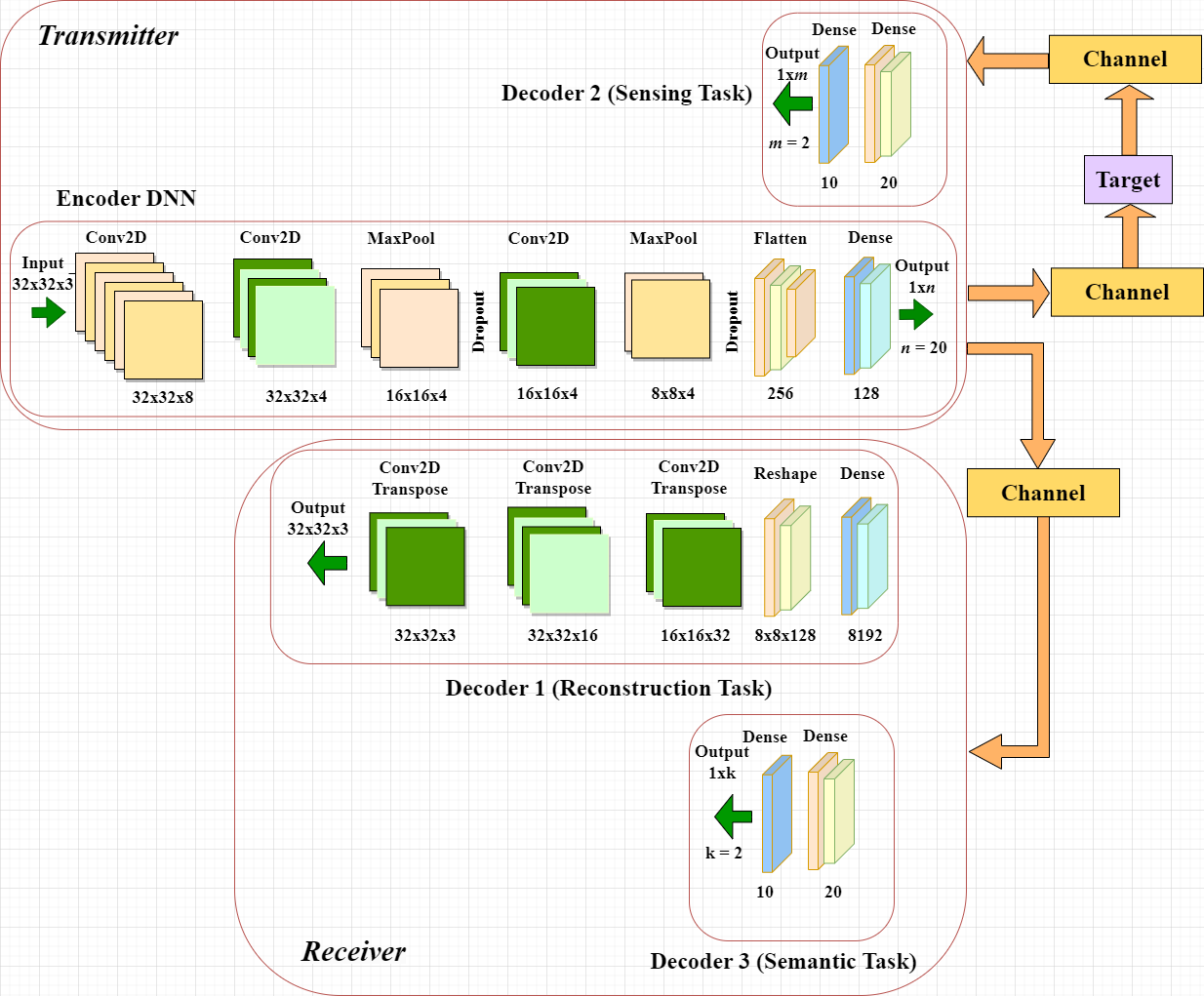}
\caption{DNN architectures for the transmitter and receiver.}
\label{fig:DNNs}
\end{figure}

The encoder and decoder architectures are parameterized by the encoder output size. The DNN architectures are selected by starting with a small number of layers and small layer sizes, and gradually increasing them until the best performance is achieved. The resulting DNN architectures are shown in Fig.~\ref{fig:DNNs}, when the encoder output size is selected as 20. A convolutional neural network (CNN) is trained for the encoder and Decoder 1, and feedforward neural networks (FNNs) are trained for Decoders 2 and 3. CNN consists of convolution layers (with the appropriate kernels and filters), maxpooling layers (for downsampling) and dropout layers (to prevent overfitting). ReLU activation is used at the hidden layers, linear activation is used at the encoder output (to generate signal for transmission over the wireless channel with constellation points mapping to continuous variables as opposed to pre-defined mappings from bits to symbols as in conventional digital modulation schemes) and the Decoder 1 output (to reconstruct input samples), and Softmax activation is used at the Decoder 2 output (for sensing decision) and Decoder 3 output (for semantic validation).

\begin{figure*}[ht!]
	\centering
	\includegraphics[width=0.325\textwidth]{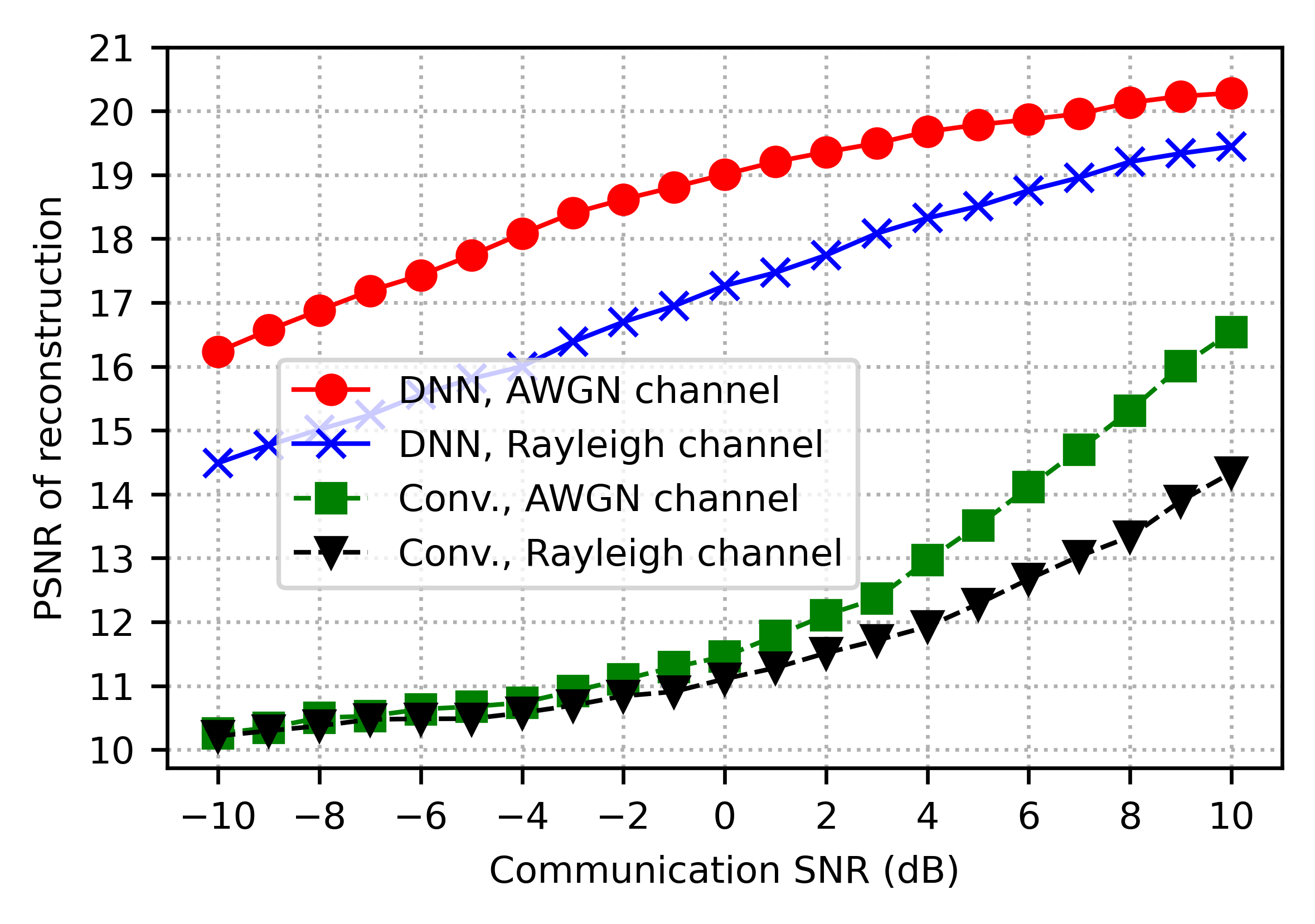}
	\includegraphics[width=0.325\textwidth]{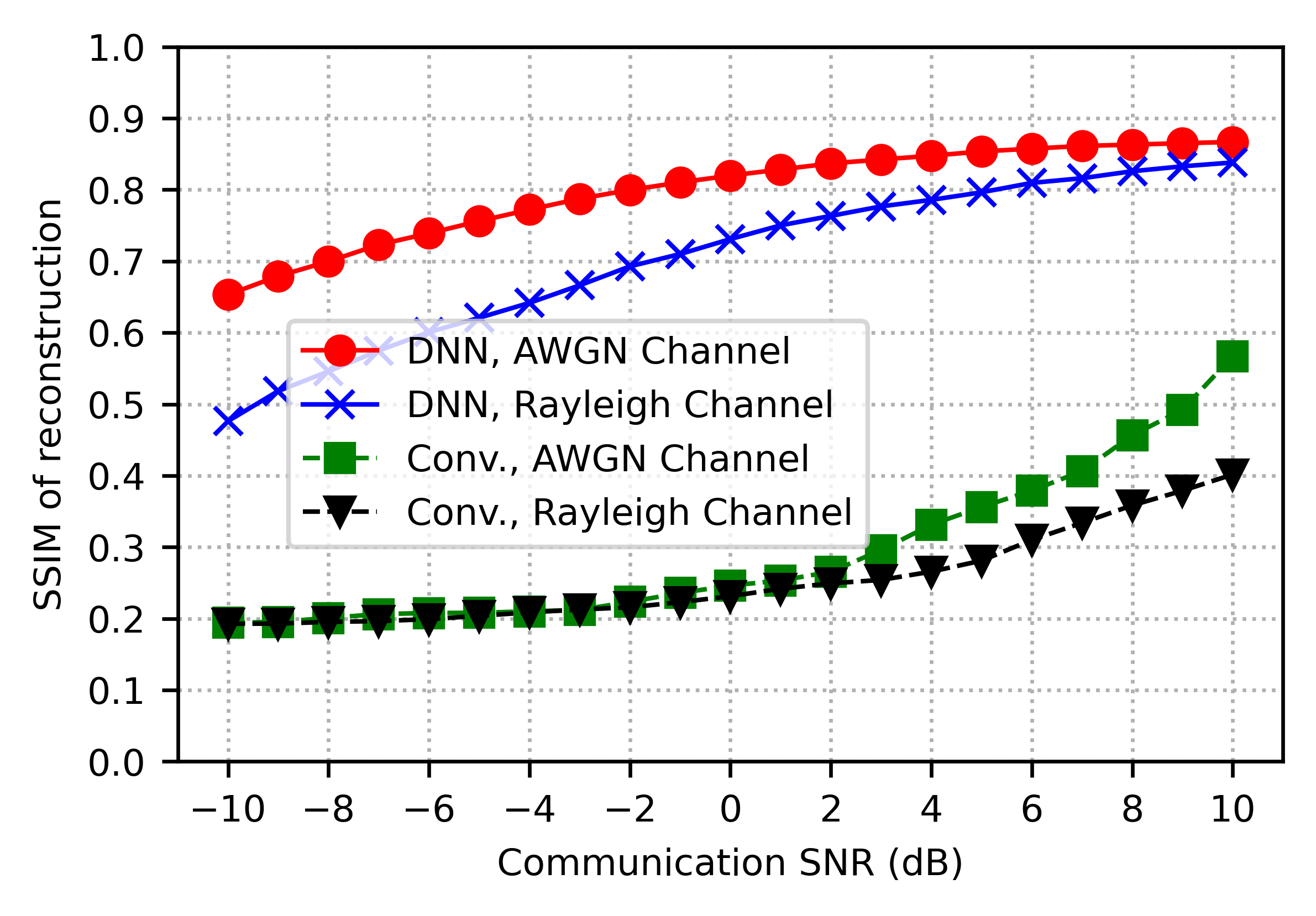}
	\includegraphics[width=0.325\textwidth]{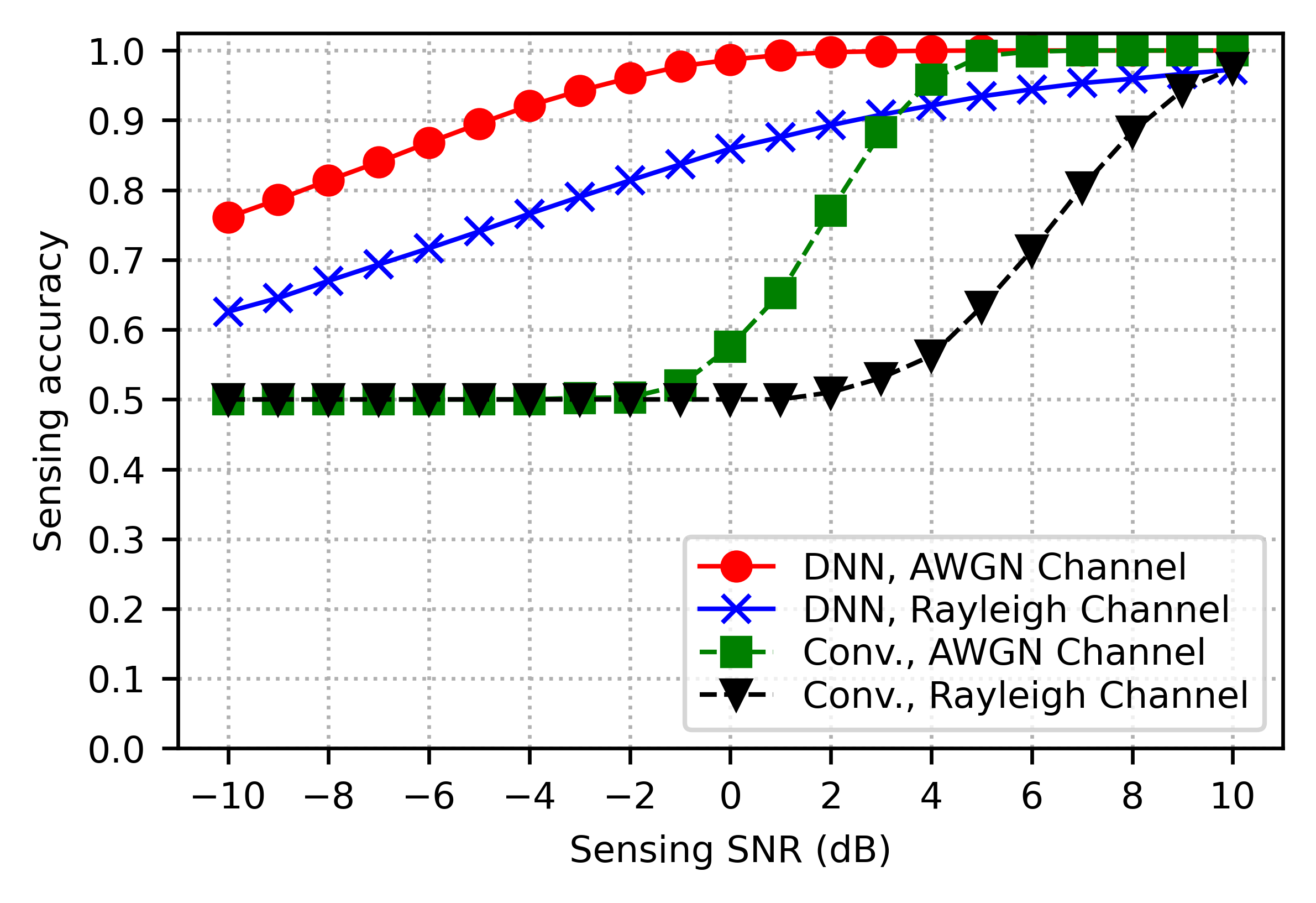}

\caption{Comparison of DNN-based schemes with conventional communications and conventional sensing (labeled as `Conv.').}
	\label{fig:JSC_compare}
 \vspace{0.2cm}
\end{figure*}

\noindent \textbf{Joint training with multi-task learning.} MTL enables joint training of the encoder and the three decoders to optimize multiple tasks (data reconstruction, sensing, and semantic validation) simultaneously, rather than training each component in isolation. We take the channel effects into account while optimizing the encoder and decoders.

For joint training, we define a combined loss function that encapsulates the objectives of each of the tasks.  A weight is assigned to each task's loss, indicating its relative importance. Then we consider the weighted sum of losses as the combined loss to ensure that the encoder and decoders work in concert to achieve their respective goals while sharing knowledge and optimizing their overall performance. Results are obtained using Python and models are trained in Keras with the TensorFlow backend, using Adam as the optimizer.

\noindent \textbf{Input data.} The CIFAR-10 dataset is used for input samples of the encoder at the transmitter. CIFAR-10 is a collection of color images from 10 classes, namely `Airplane',	`Automobile', `Bird', `Cat', `Deer', `Dog', `Frog', `Horse', `Ship', `Truck'. The size of each data sample (image) is $32 \times32 \times 3$. Each data sample consists of red, green, blue (RGB) pixels. A pixel in each RGB component takes a value between $0$ and $255$. We normalize these pixel values to the range between 0 and 1 before feeding them to the encoder DNN at the transmitter. We consider 50,000 training samples and 10,000 test samples for numerical results. 

\noindent \textbf{Channel effects.} The encoder output is sent over the discrete-time complex-valued wireless channel. We consider the following channel effects that are learned by joint training of the encoder and the encoders:
\begin{itemize}
    \item AWGN channel, where the signal received at the receiver is the transmitted signal plus the white Gaussian noise. 
    \item Rayleigh fading channel, where the transmitted signal is subject to Rayleigh fading (the real and imaginary parts of the complex Rayleigh fading channel is chosen as independent Gaussian variables) and white Gaussian noise is added to the received signal. 
\end{itemize}
The overall model is augmented with Gaussian noise layer and Rayleigh fading channel layer with the appropriate SNR added between the encoder and each decoder. The channels between the transmitter and the receiver, and the channels between the transmitter and the target undergo different channel effects by accounting for noise and fading. In training, a channel realization is randomly drawn from the given channel distribution for each data sample. In test time, the channel distribution remains the same even though individual realizations vary and are not known in advance to the already trained DNN models.

For sensing, the overall channel is a concatenation of channel from the transmitter to the target and from the target back to the transmitter. We consider only fading and noise effects in this paper while the channel frequency shift (due to the relative speed of potential target) and phase shift (due to the reflection off the target) can be further incorporated to enrich the sensing environment.

\noindent \textbf{Performance measures.}  
The performance measures for the three underlying tasks of communications, sensing, and semantic information validation are given by: 
\begin{itemize}
    \item \emph{PSNR}: a measurement of the quality of a signal (image), by evaluating the ratio between the maximum possible signal value and the noise level determined by the MSE where the input and reconstructed pixels are compared to each other.
    \item\emph{SSIM}: similarity measure of two images with taking the luminance, contrast and the structure into account.
    \item \emph{Sensing accuracy}: average accuracy of signal classification task for sensing performed by Decoder 2 at the transmitter (signal classification task can either detect only whether the target is present or not, or further incorporate decisions on potential ranges of the target).
    \item \emph{Semantic accuracy}: the average accuracy of the classification task performed by Decoder 3 at the receiver to validate whether the semantic information is preserved during the information transfer, i.e., whether the underlying classification task can correctly identify the output labels of data samples, e.g., from 10 classes of CIFAR-10 dataset, or which label subset of interest they belong to, e.g., the image contains an animal.  
\end{itemize}

\noindent \textbf{System parameters.} The system model involves several tunable parameters. We will vary these parameters to measure the performance. 
\begin{itemize}
\item \emph{Multi-task weights}: weights used to aggregate the losses of different tasks in the combined loss function of MTL, namely sensing and communications in addition to the extension to semantic communications.   
\item \emph{Communication SNR}: SNR of signals received as the input of decoders at the receiver (Decoder 1 and Decoder 3). 
\item \emph{Sensing SNR}: SNR of signals reflected by a potential target and received as input to the decoder at the transmitter (Decoder 2). 
\item \emph{Encoder output size}: number of symbols transmitted as output of the encoder at the transmitter (represents the compression of the encoder with respect to the size of input samples and determines the input size of all three decoders). The output size of the encoder DNN corresponds to the block length of the transmitted complex codeword where the in-phase and quadrature components are concatenated. As a result, $n$ symbols at the encoder output correspond to $n/2$ complex symbols.
\item \emph{Number of potential target ranges}: number of potential ranges for the positions where a potential target may be located (this number determines the number of labels in the classification problem for sensing since the sensing objective is to identify the presence and if so, the range of a target). 
\end{itemize}

\noindent \textbf{Baselines.}  
We compare the performance of JSC with  
four baselines of conventional communications, conventional sensing, individual sensing and individual communications.
\begin{itemize}
    \item \emph{Conventional communications}: Separate source and channel coding modules are used with JPEG-2000 and Reed-Solomon error correction algorithms, respectively and 16-QAM is the modulation scheme.
    \item \emph{Conventional sensing}: An energy detector is used at the transmitter which compares the reflected signal power with a threshold. The overall channel is a concatenation of channel from the transmitter to the target and from the target back to the transmitter.
    \item \emph{Individual sensing}: The encoder and the decoder DNNs are trained for the sole purpose of sensing such that the sensing loss function is the only one assigned a non-zero weight in MTL.
    \item \emph{Individual communications}: The encoder and the decoder DNNs are trained for the sole purpose of communications such that the reconstruction loss function is the only one assigned a non-zero weight in MTL.
\end{itemize}

\section{Performance of Joint Sensing and Communications} \label{sec:perf}

We consider the following default values of system parameters for performance evaluation. In MTL, the weight for reconstruction loss is 0.95, the weight is 0.05 for sensing loss in JSC, and the weights are 0.025 for sensing and semantic task losses in JSSC. Communication SNR is 3dB. Sensing SNR is 3dB. Size of encoder output is 20. Number of potential target ranges is 1. For performance evaluation, we vary system parameters one at a time by setting the rest of parameters to default values.

\noindent \textbf{Comparison with conventional communications and sensing.} 
In conventional communications, the maximum compression ratio from original CIFAR image to JPEG-2000 is 20 (3072 bytes to 152 bytes), resulting in 510 complex symbols with Reed-Solomon codes and 16-QAM modulation, maintaining a coding rate of 152/255. Note that the DNN-based approach can achieve much higher compression at the encoder output. PSNR and SSIM values are assessed across -10dB to 10dB communication SNR with 1 dB steps, acknowledging the sensitivity of decoding received JPEG-2000 files to bit errors, especially in low SNR regimes. For comparison, we consider single-task learning of DNN-based sensing or communications by setting the weight of one task to 1 and the other one to 0. Fig.~\ref{fig:JSC_compare} shows that, in both AWGN and Rayleigh channels, the DNN-based approach surpasses the conventional  approaches in terms of PSNR, SSIM, and sensing accuracy across all SNRs.

\begin{figure*}[t!]
	\centering
 \includegraphics[width=0.325\textwidth]{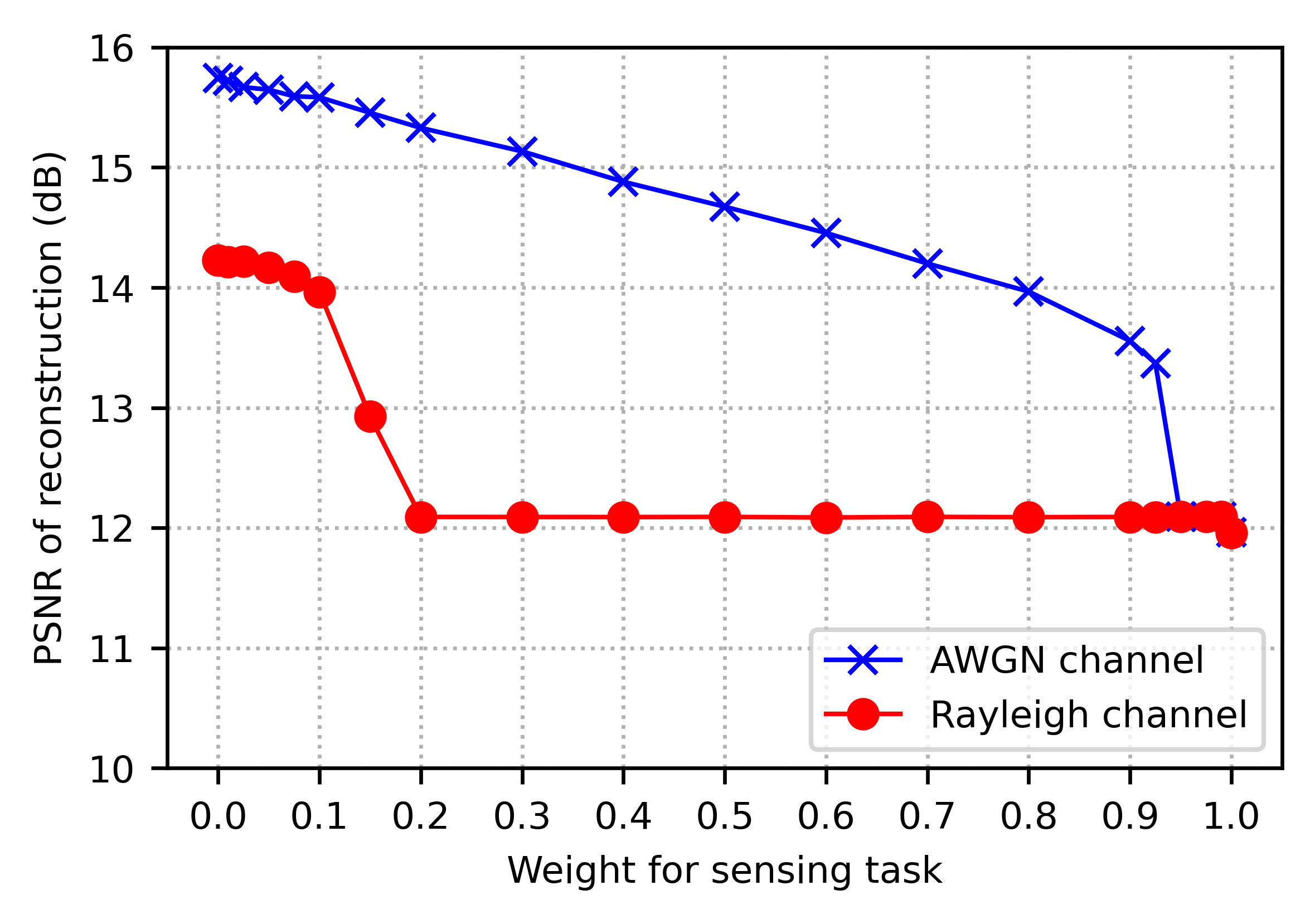}
  \includegraphics[width=0.325\textwidth]{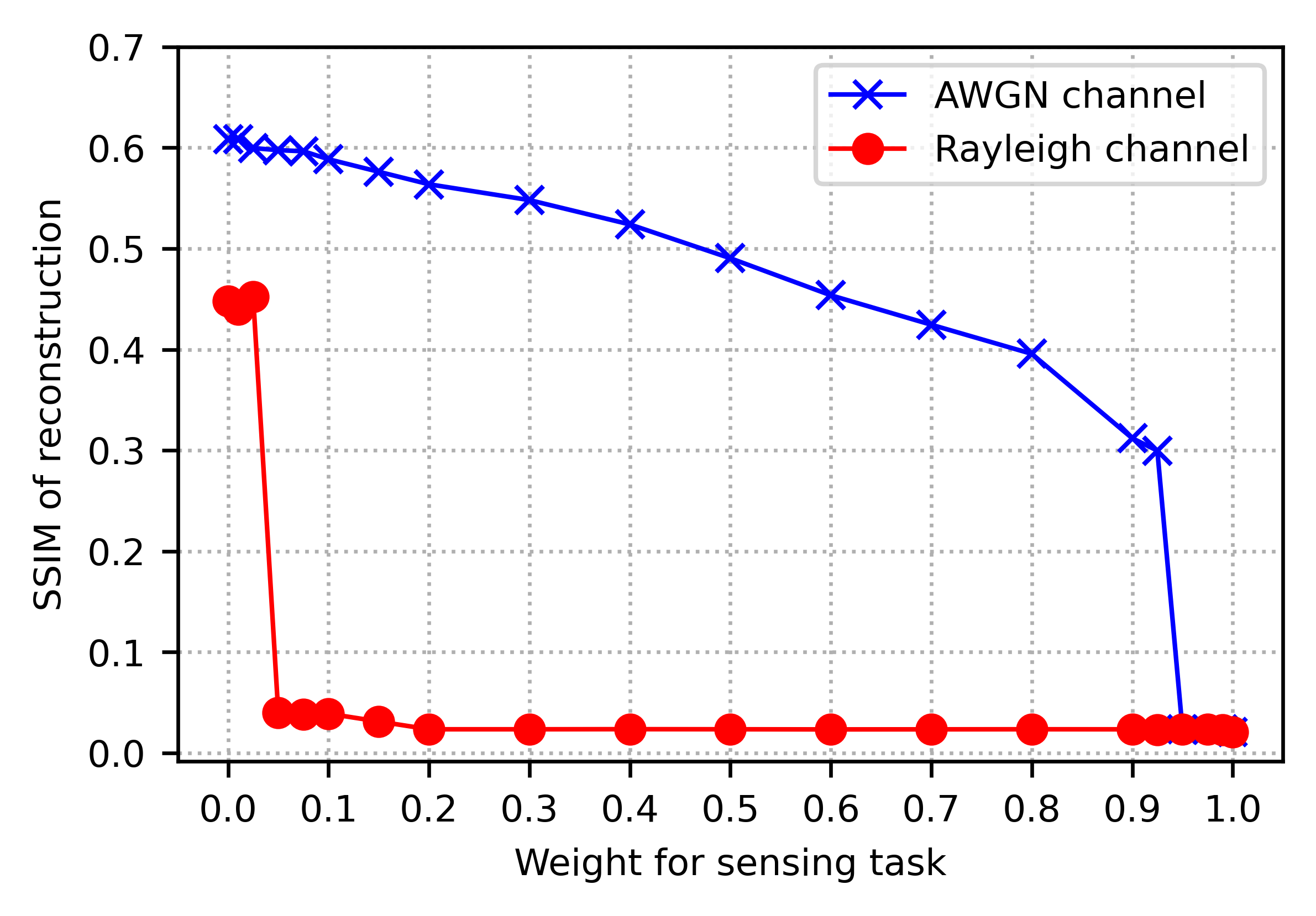}
   \includegraphics[width=0.325\textwidth]{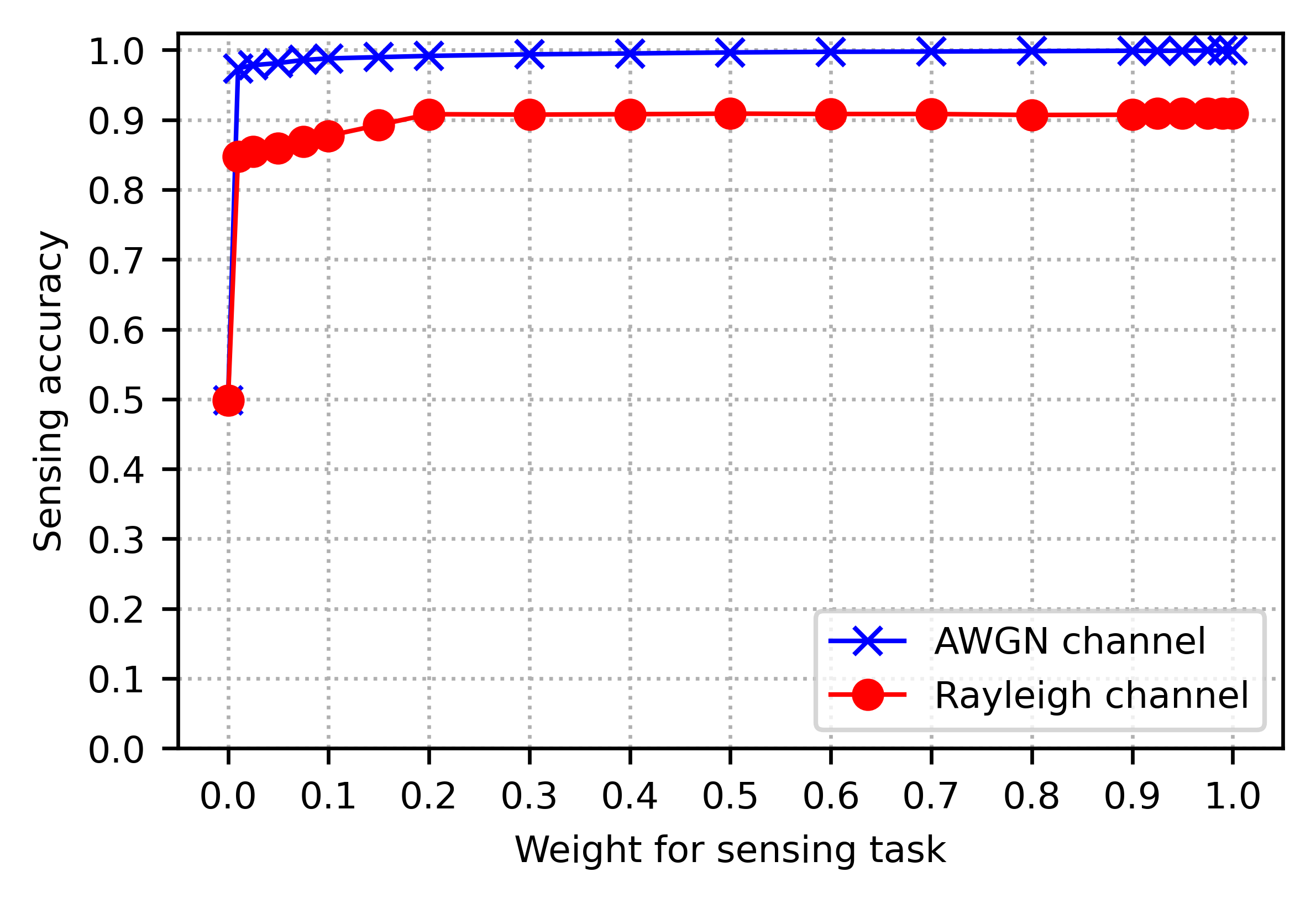}
	\includegraphics[width=0.325\textwidth]{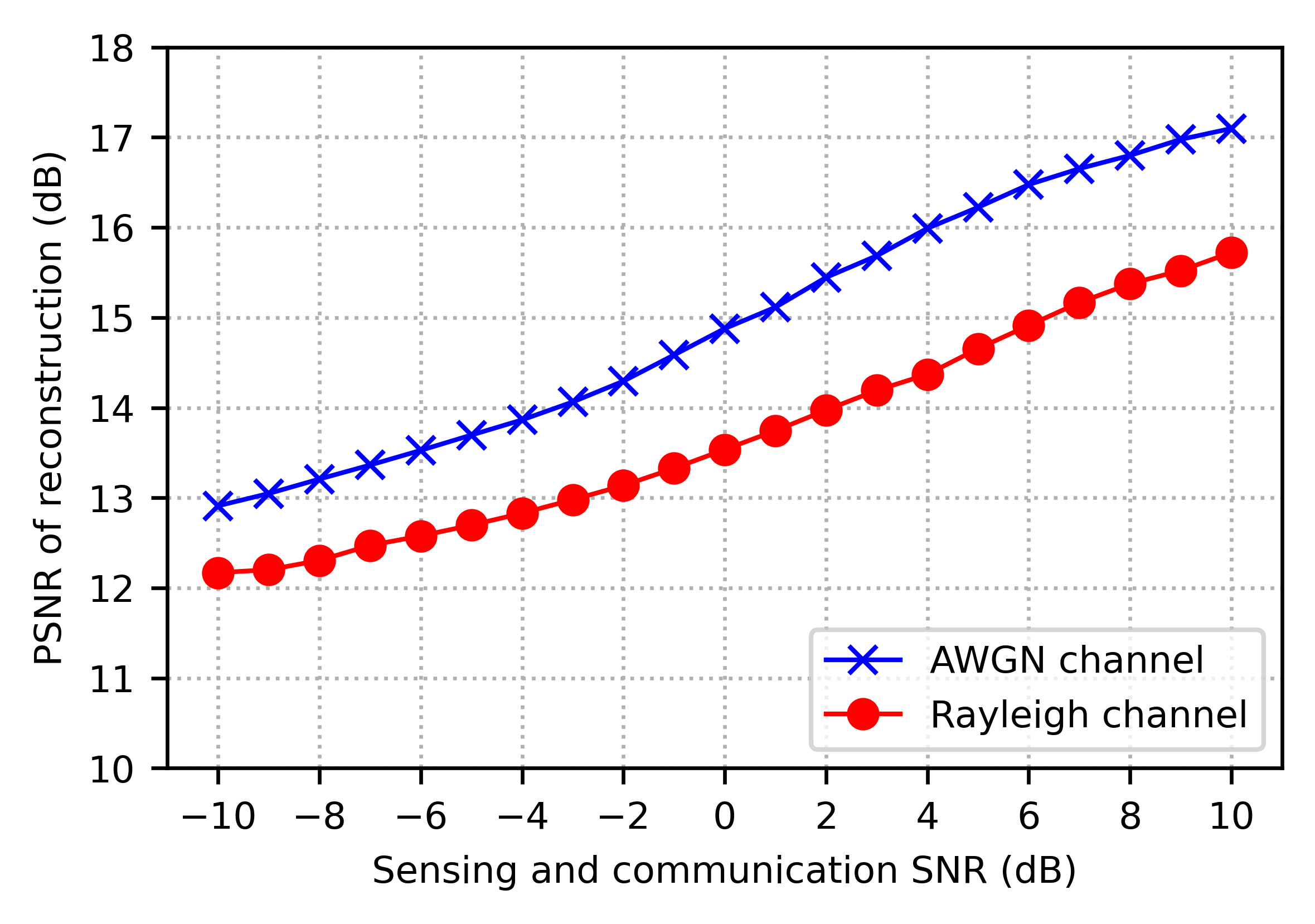}
 	\includegraphics[width=0.325\textwidth]{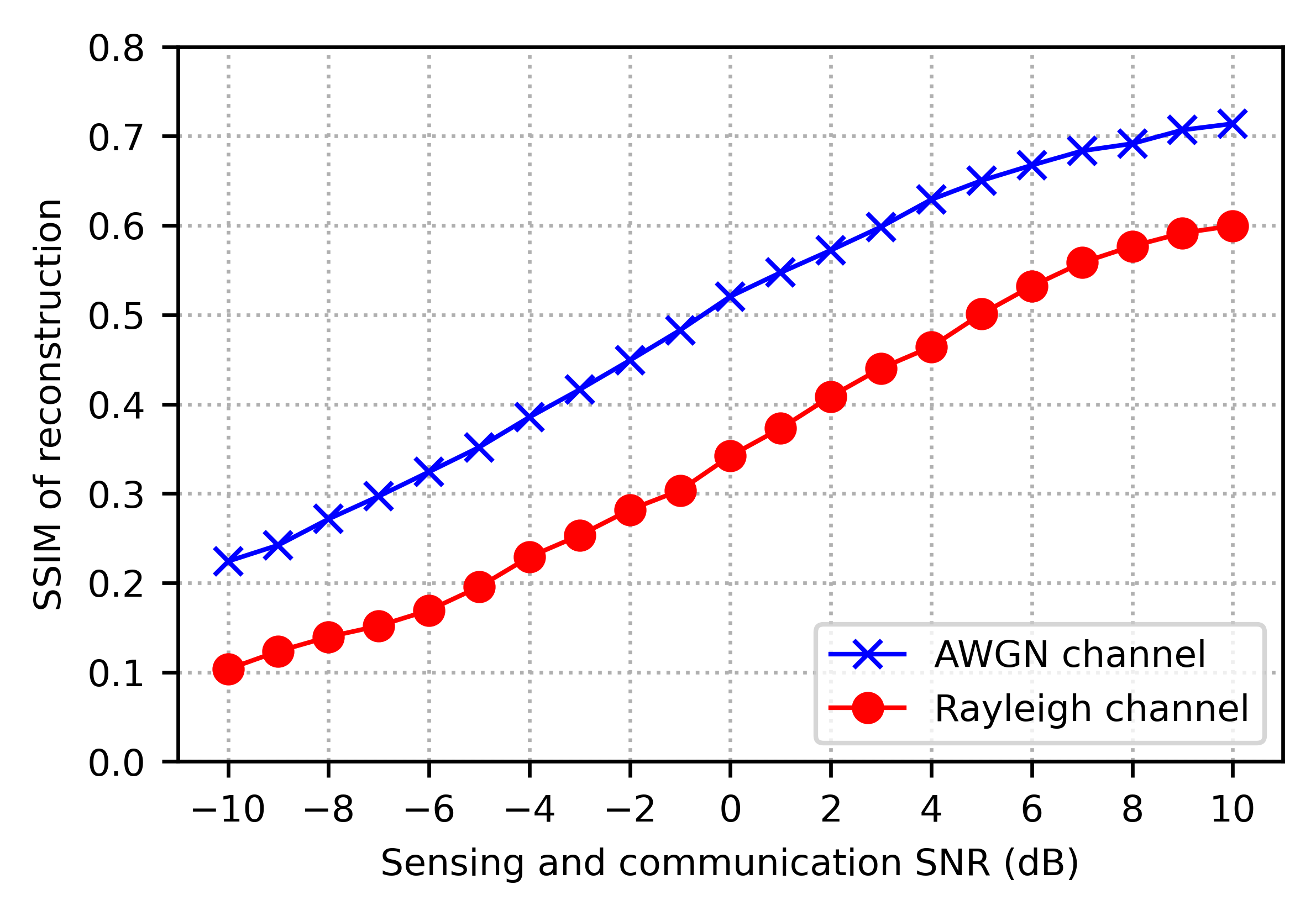}
  	\includegraphics[width=0.325\textwidth]{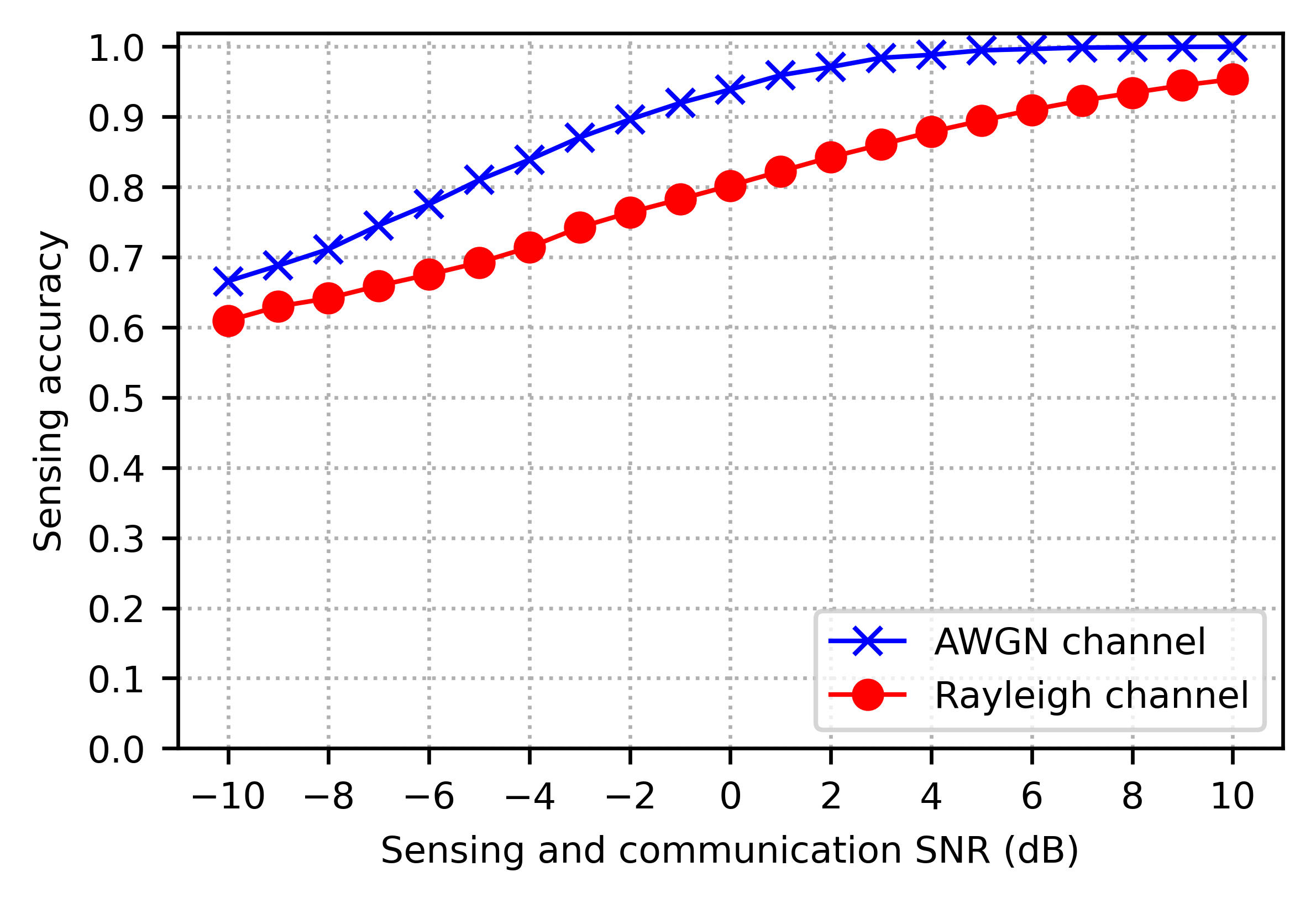} 
	\includegraphics[width=0.325\textwidth]{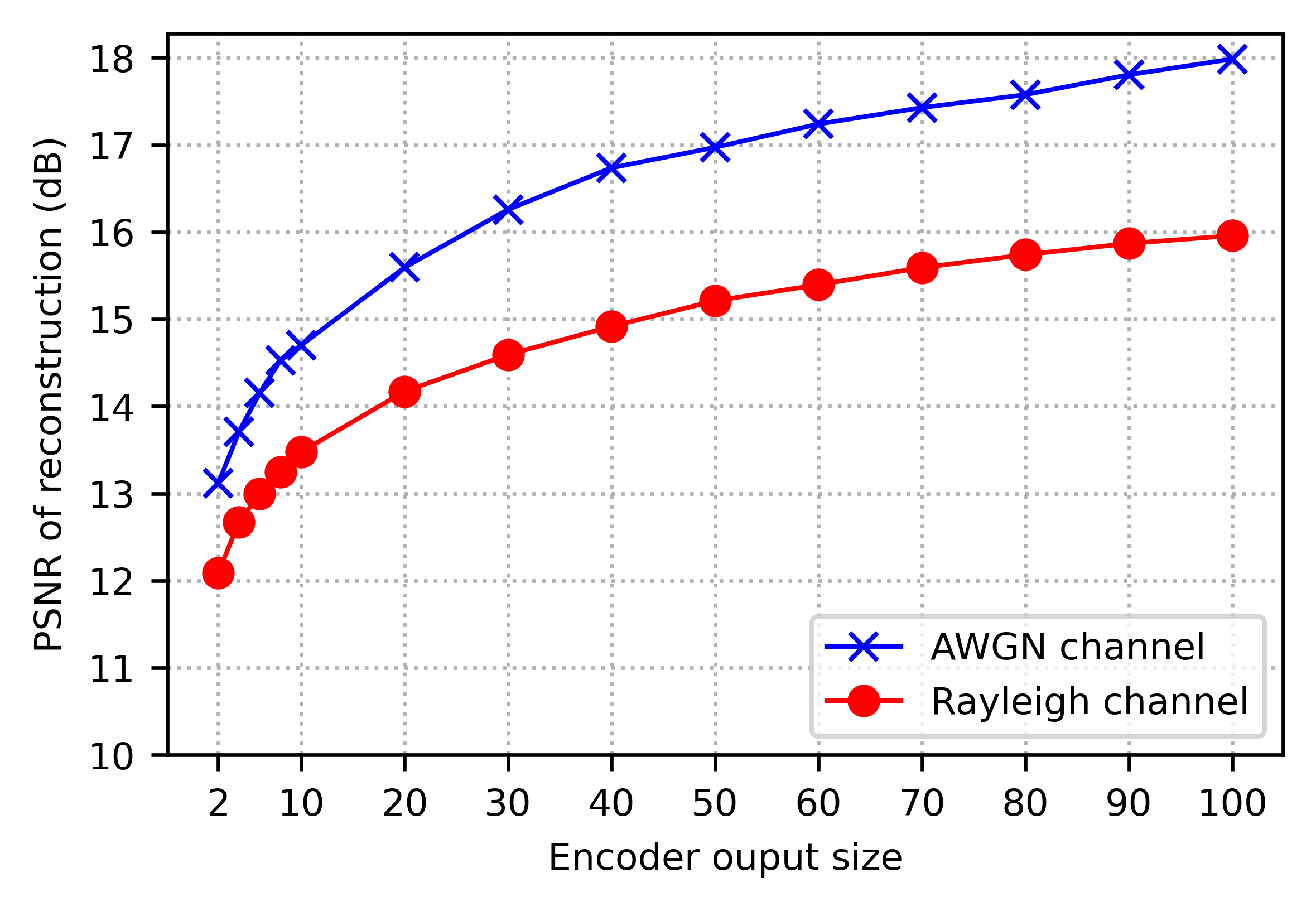}
 	\includegraphics[width=0.325\textwidth]{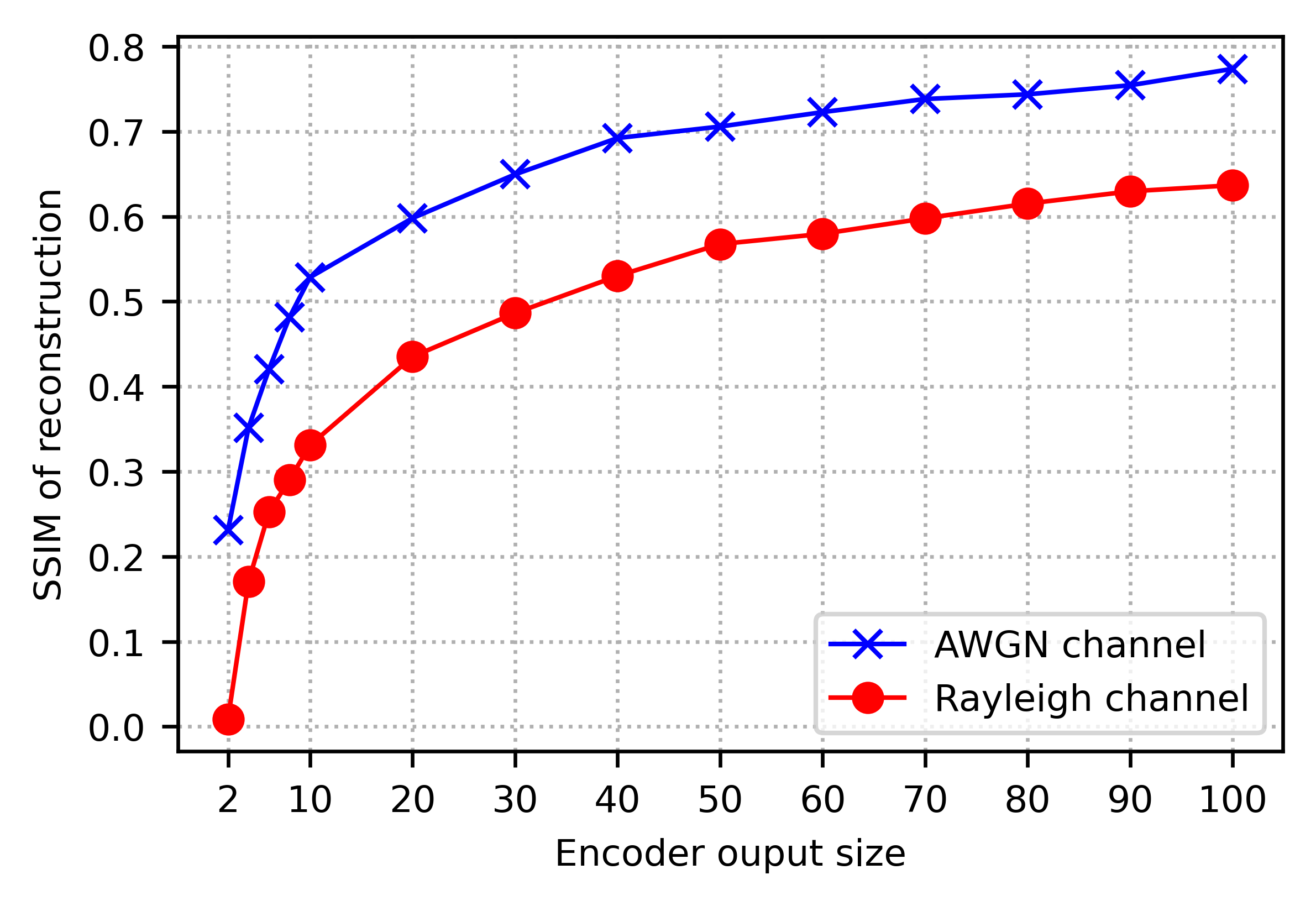}
  	\includegraphics[width=0.325\textwidth]{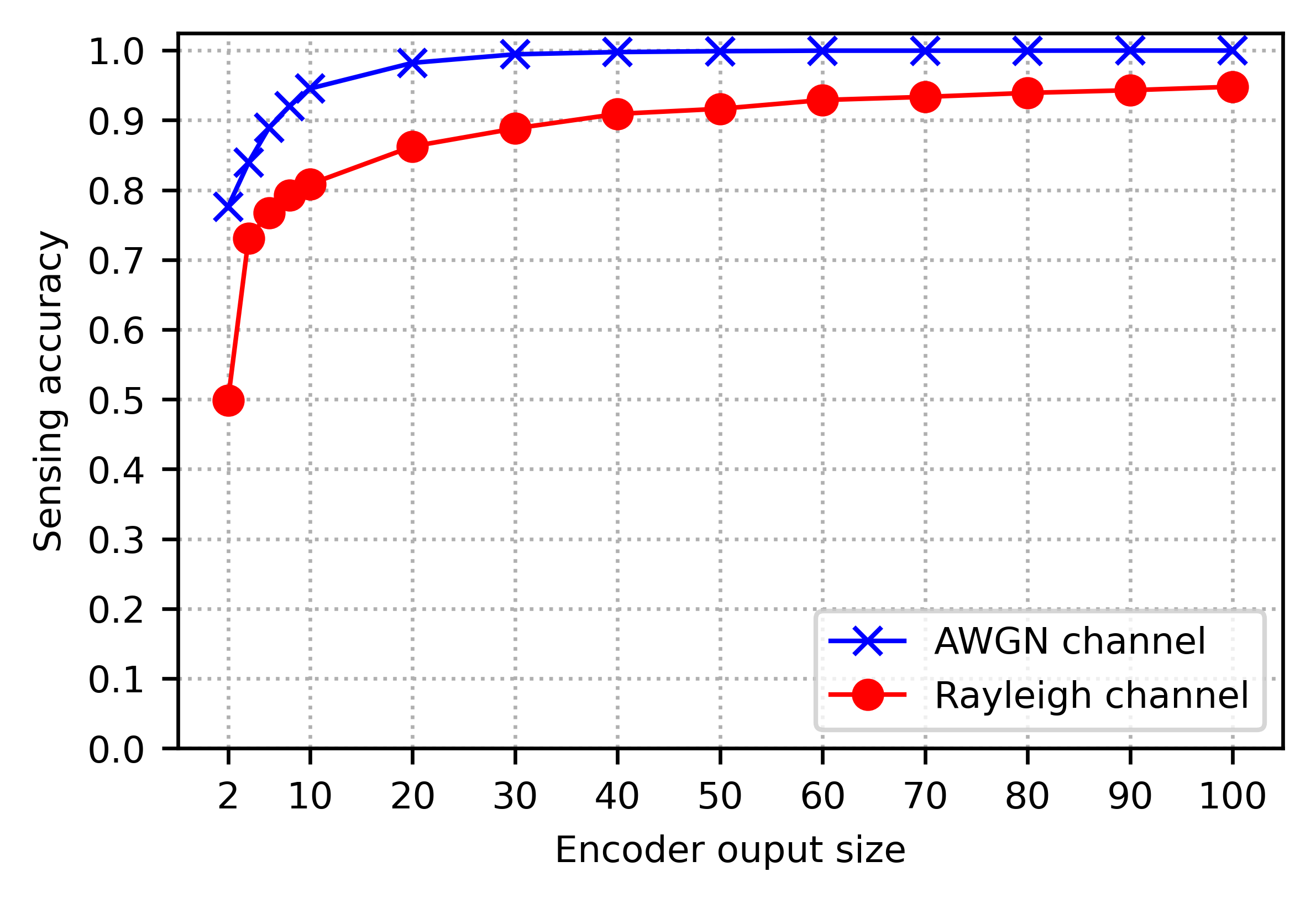}
	\includegraphics[width=0.325\textwidth]{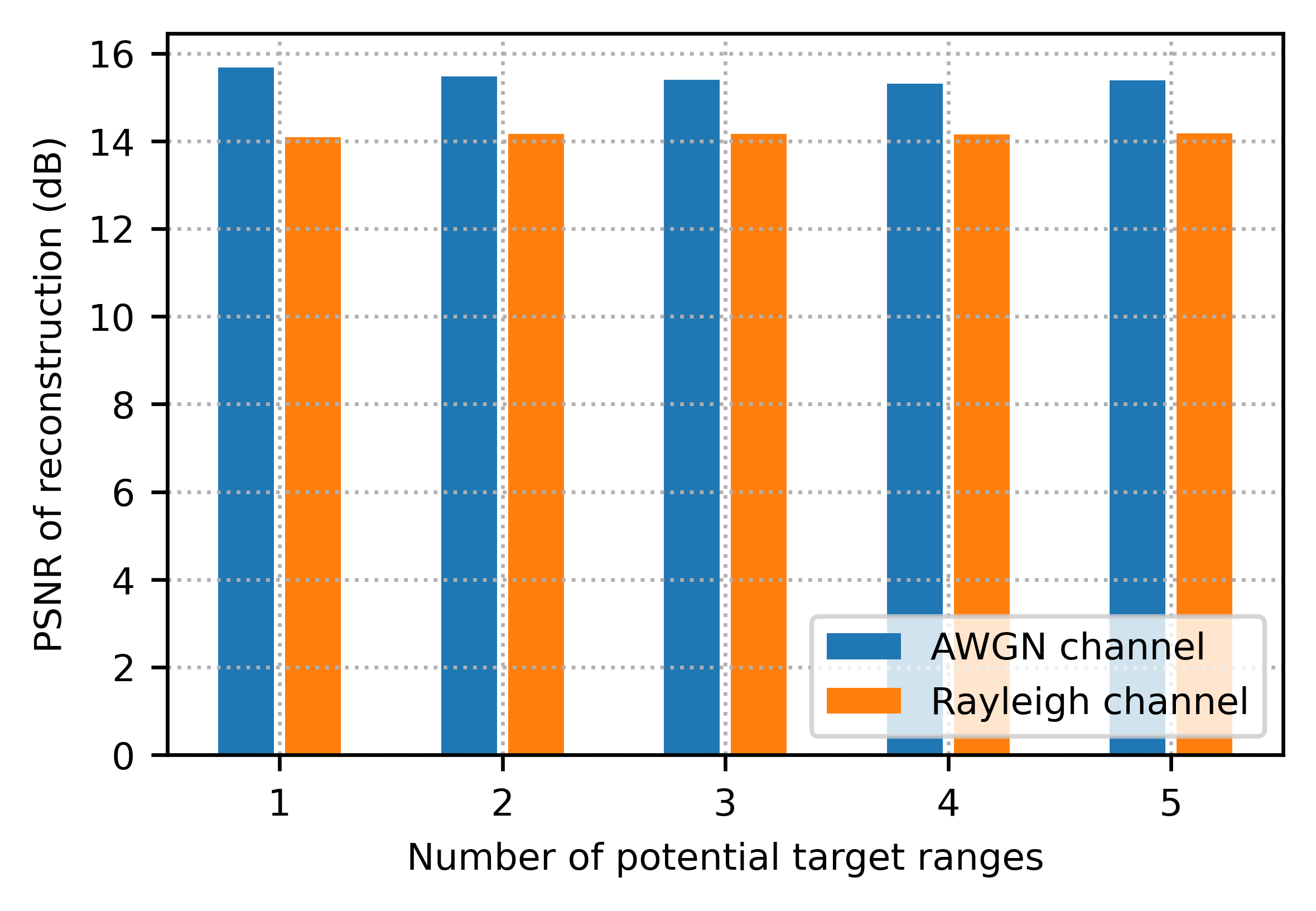}
 \includegraphics[width=0.325\textwidth]{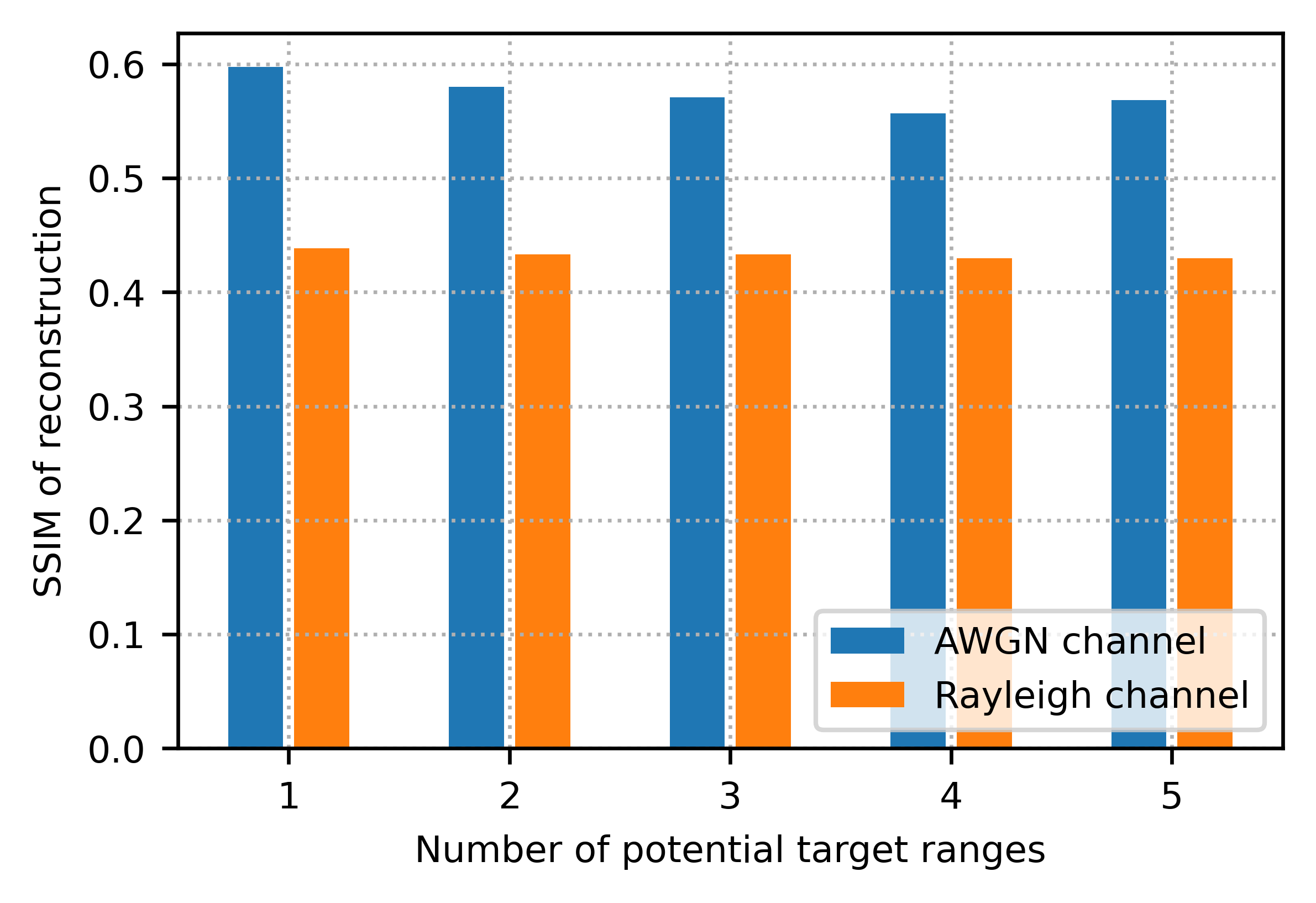}
 \includegraphics[width=0.325\textwidth]{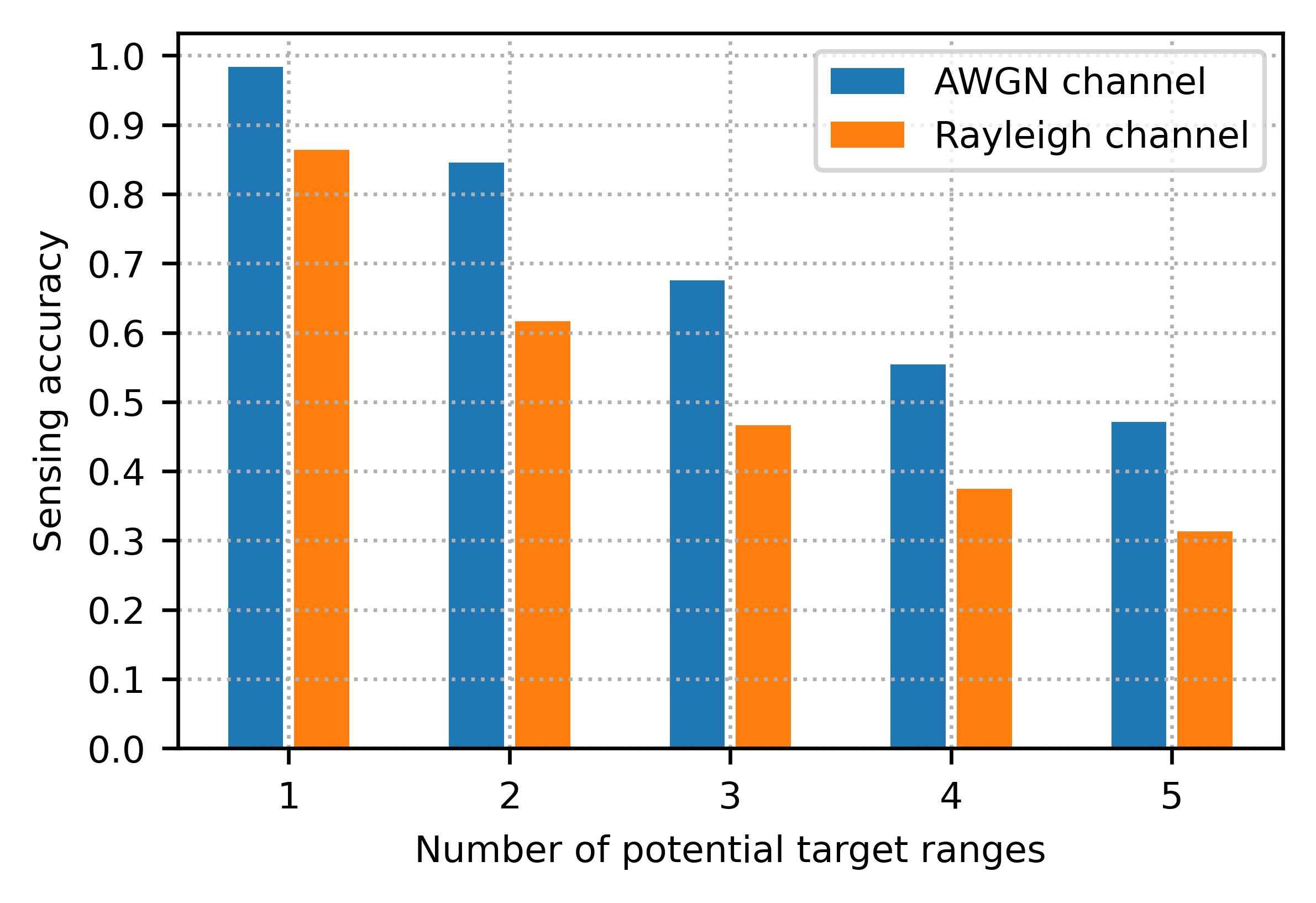}
 \caption{JSC performance as a function of system parameters (multi-task weights, SNR, encoder output size, and number of potential target ranges).}
\label{fig:restpar}
\end{figure*}

\noindent \textbf{Joint vs. individual sensing and communications.} We vary the weight of sensing loss for MTL to highlight the value of MTL over single-task learning of sensing or communications. PSNR, SSIM, and sensing accuracy are shown in Fig.~\ref{fig:restpar} for both AWGN and Rayleigh channels. These results show the value of MTL for JSC in contrast to individual optimization of either sensing or communications. When the sensing task weight is zero, we end up with the baseline of individual communications such that reconstruction is highly successful (reconstruction loss is low) but sensing is not successful (sensing accuracy is very low). When this sensing task weight is 1, we end up with the baseline of individual sensing such that sensing is highly successful (sensing accuracy is high) but reconstruction is not successful (reconstruction loss is very high). When select the sensing weight in between, we can achieve a desirable performance for both sensing and reconstruction, i.e., the sensing accuracy is high while the reconstruction loss remains low. 

\begin{figure*}[t!]
	\centering
	\includegraphics[width=0.425\textwidth]{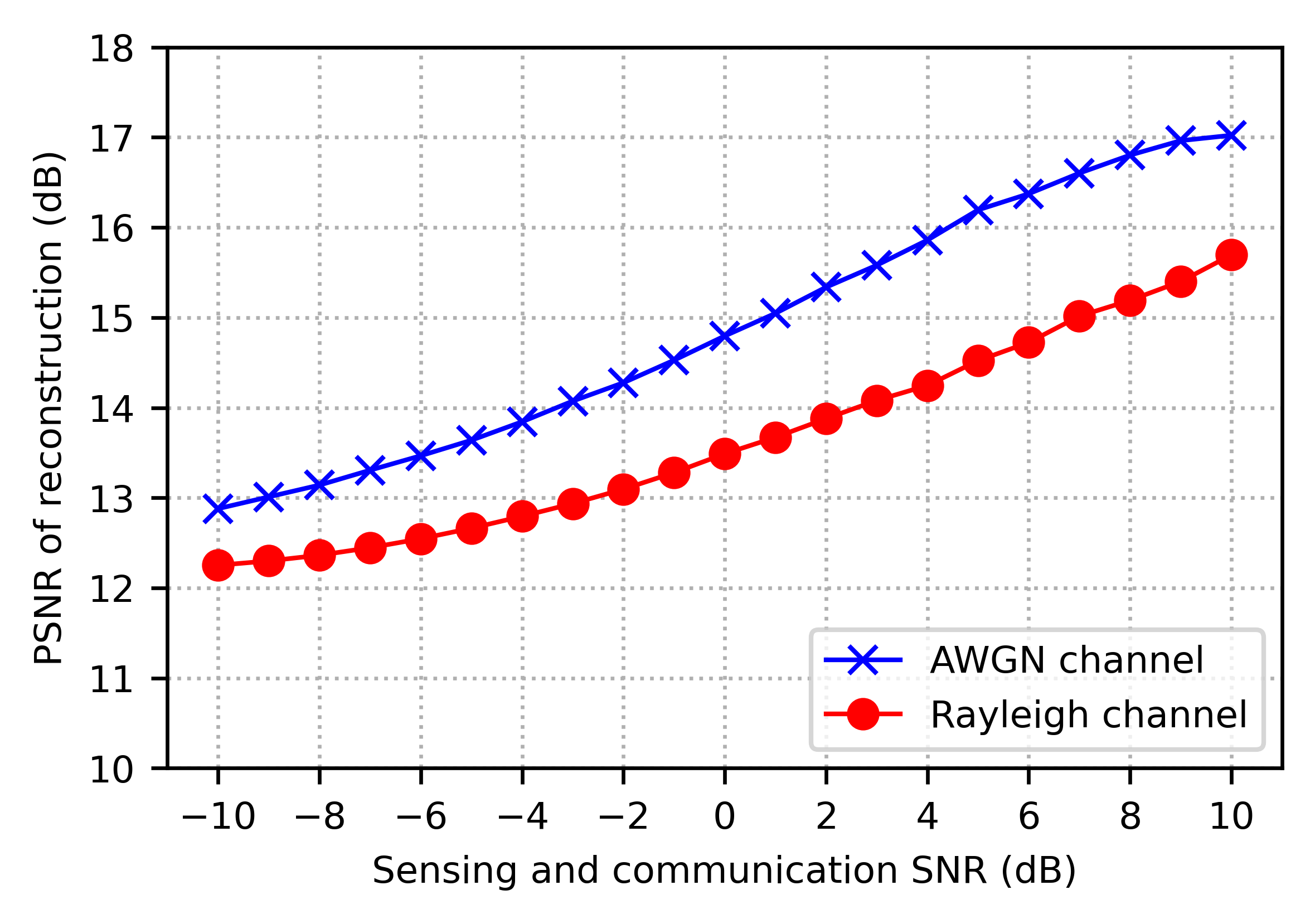} \hspace{1cm}
	\includegraphics[width=0.425\textwidth]{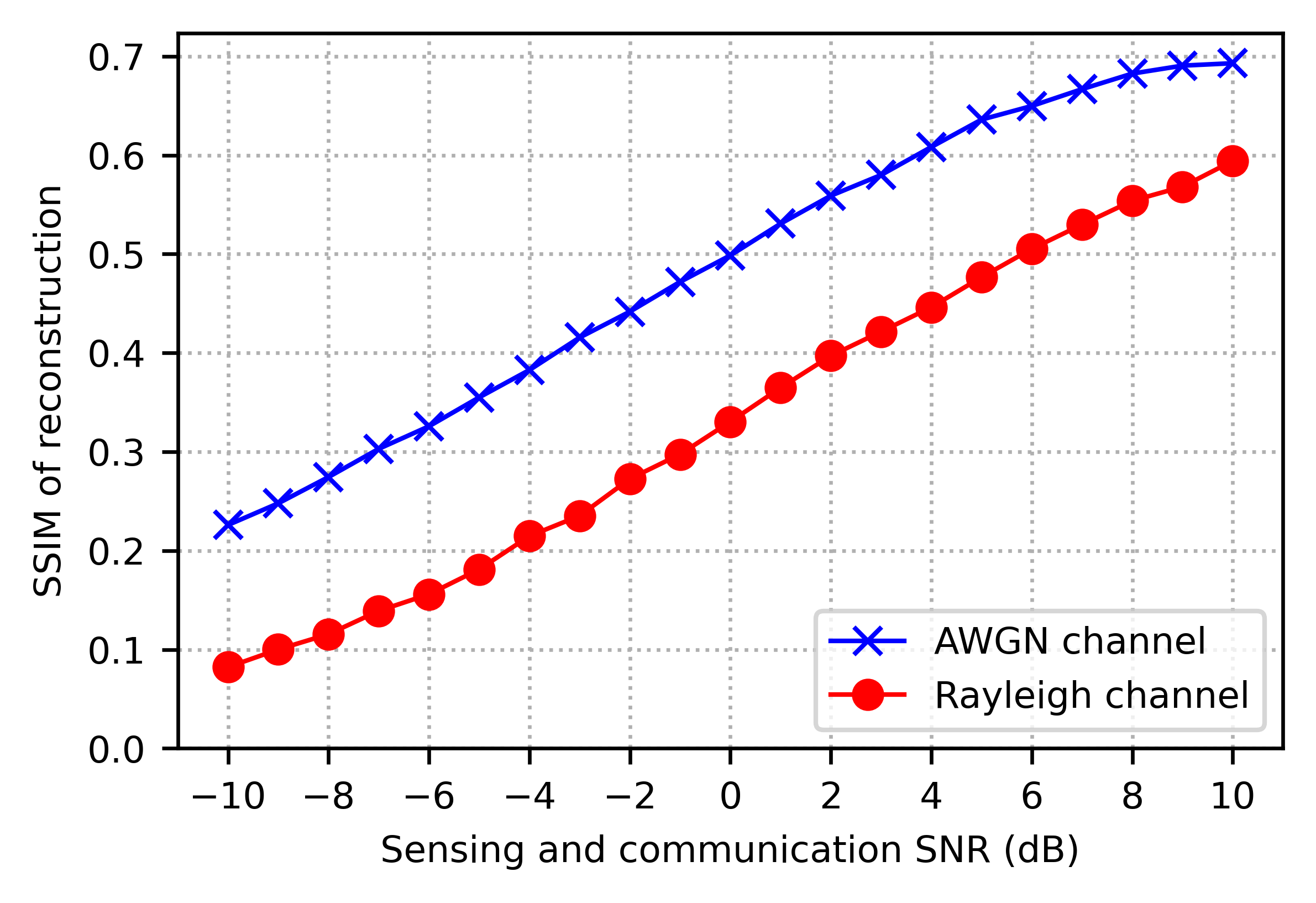}
	\includegraphics[width=0.425\textwidth]{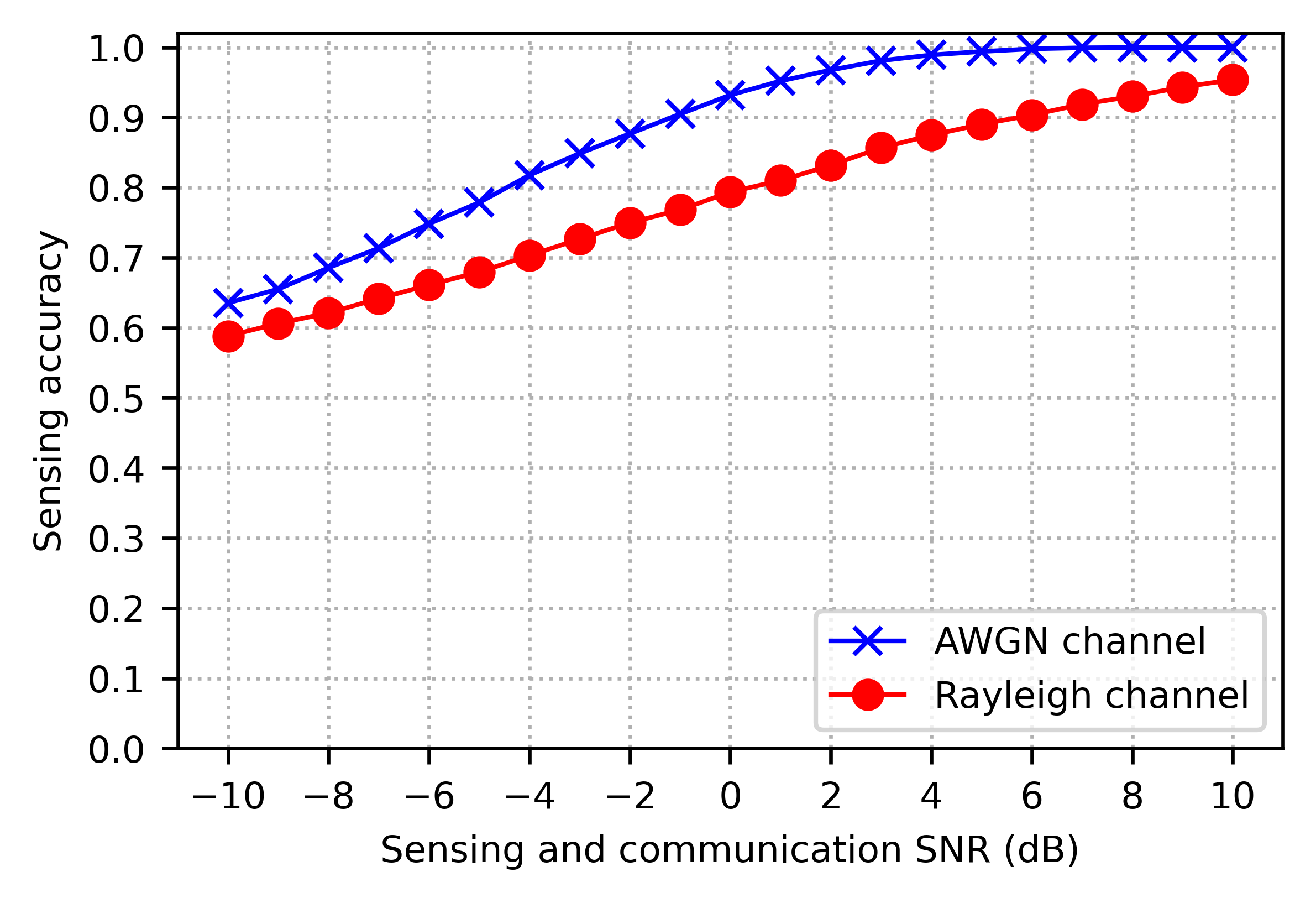} \hspace{1cm}
	\includegraphics[width=0.425\textwidth]{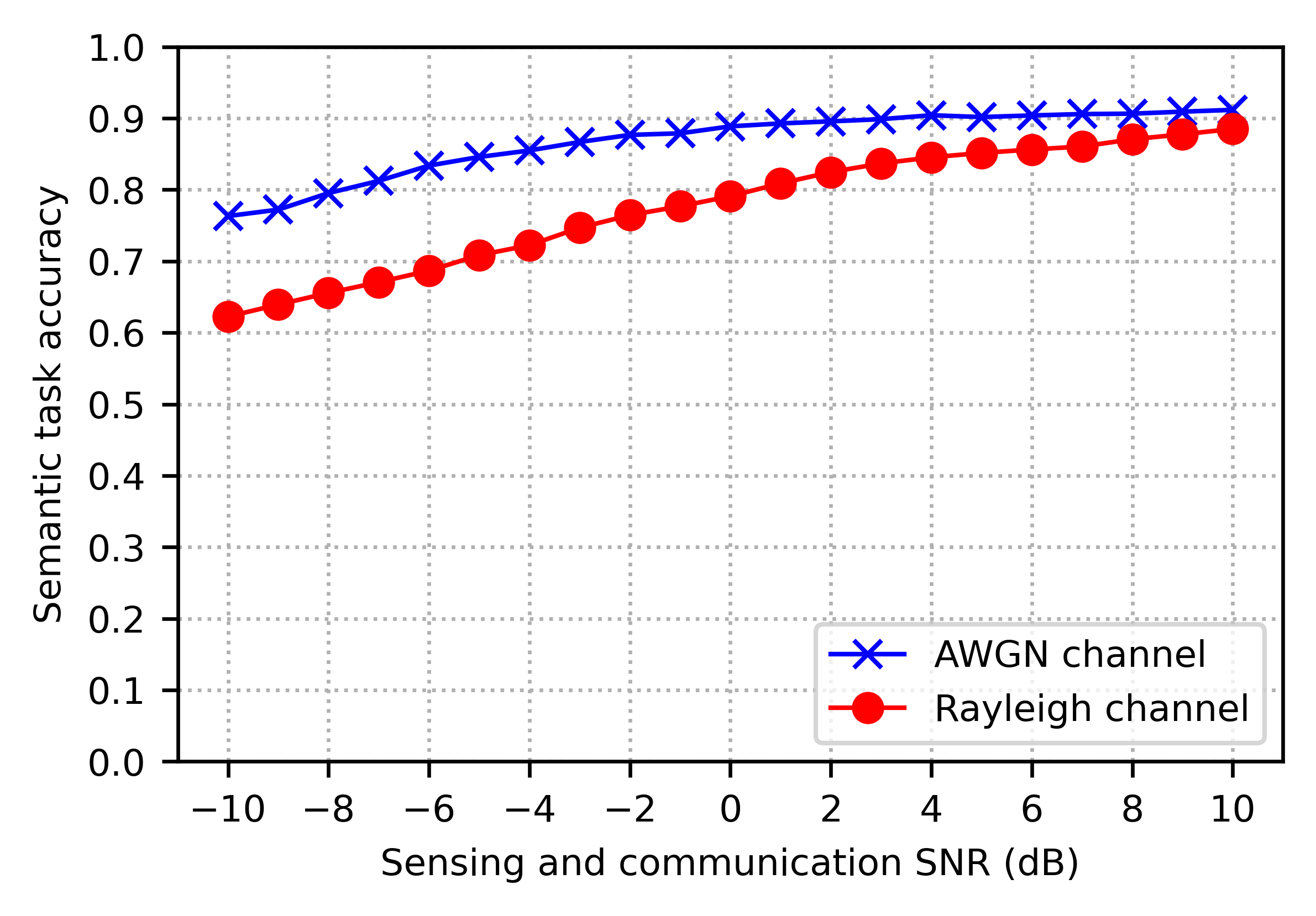}
\caption{Joint sensing and semantic communication performance vs. SNR (when sensing SNR is the same as communication SNR, both measured in dB).}
	\label{fig:JSC_semantic}
 \vspace{0.2cm}
\end{figure*}

\noindent \textbf{Effects of system parameters on JSC}. Next, we vary the rest of system parameters and show the performance in Fig.~\ref{fig:restpar}. First, we evaluate sensing accuracy, PSNR and SSIM when we vary the communication SNR to measure its effect on PSNR and SSIM performance and separately vary the sensing SNR to measure its effect on sensing performance. Results in Fig.~\ref{fig:restpar} show that the performance quickly improves with the SNR for both AWGN and Rayleigh channels. In all cases, the learning performance is better under the AWGN channel compared to the more difficult case of Rayleigh channel, as we will observe also for other performance results.

Next, we show the performance as a function of encoder output size at the transmitter. The smaller this size becomes, the more efficiently the underlying resources such as channel access and transmit energy are utilized potentially with less delay. Results in Fig.~\ref{fig:restpar} show that the performance quickly improves for both AWGN and Rayleigh channels when we start increasing the encoder output size. JSC can achieve high sensing accuracy, PSNR, and SSIM with high resource efficiency, e.g., when the encoder output size is selected as 20, i.e., the dimension is reduced from 3072 (namely, 32 $\times$ 32 $\times$ 3) for input samples to 20 in latent space. 

Next, we extend the sensing problem from identifying whether a target is present or not, to classifying the range (proximity) of the target (including the absence of the target). The sensing SNR (with default value of 3dB) corresponds to the closest potential position, whereas the SNRs for other ranges are made incrementally smaller. Results in Fig.~\ref{fig:restpar} show the performance as a function of the number of potential target ranges. As we add more potential ranges for the target, the number of labels increases along with the difficulty of the underlying multi-label classification problem such that the performance starts dropping (especially observed for sensing accuracy under Rayleigh channel).

\noindent \textbf{Joint sensing and semantic communications.}
Next, we add semantic communications to the MTL framework of JSC, and jointly optimize the objectives of three tasks, namely sensing, information reconstruction, and semantic information validation. Decoder 3 performs the semantic information validation task at the receiver. For the CIFAR-10 data input, we can check different levels of semantic information preserved in the transferred information such as classification labels themselves or their membership in different sets of labels. For numerical results, we consider semantic validation in terms of checking whether the transferred image belongs to a certain set of labels of interest, specifically, whether the image includes an animal (i.e., the label is `Bird', `Cat', `Deer', `Dog', `Frog', or `Horse'), or not.  We measure the performance by varying the SNR for both communications and sensing while keeping the rest of system parameters at their default values. Fig.~\ref{fig:JSC_semantic} shows that the PSNR, SSIM, sensing accuracy, and semantic task accuracy (for semantic information validation) can quickly increase with the SNR for both AWGN and Rayleigh channels. The implication is that it is possible to send syntactic information over the wireless channel while preserving the semantic information and sense the target, by using the same signal for communications and sensing.

\section{Conclusion and Future Research Directions} \label{sec:Conclusion}
We have presented the MTL framework of deep learning for JSC, introducing the addition of semantic communication capabilities. The key components of the system are the dual-purpose encoder for sensing and communications, and decoders for data reconstruction, sensing, and semantic information validation. By harnessing the power of MTL, these components collectively contribute to optimizing target sensing and data transfer while preserving semantic information, and offer a versatile, resource-efficient, and high-fidelity solution, capable of addressing the demands of real-time decision-making, context-aware data sharing, and efficient resource utilization. 

Future research directions that can build upon the MTL framework of JSSC, include working with other data modalities such as text, in addition to images, and extending the setting from point-to-point networks. In the latter, multiple semantic tasks can be incorporated for different receivers. In terms of sensing, additional objectives such as target direction identification and target tracking can be added to MTL. Finally, tasks can be enriched by adding security objectives such as anti-jamming, covert communications and protection against adversarial machine learning attacks that aim to fool the DDNs underlying the JSSC process into making errors in MTL. 

\bibliographystyle{IEEEtran}
\bibliography{arxivreferences}

\end{document}